\documentclass[aps,prl,twocolumn,superscriptaddress,groupedaddress,longbibliography,showpacs]{revtex4-1}  

\usepackage{graphicx}  
\usepackage{bm}        
\usepackage{amssymb}   
\usepackage{amsmath}
\usepackage{amsfonts}
\usepackage{natbib}
\usepackage{float}
\usepackage{xcolor}
\usepackage{ulem}
\usepackage{txfonts}
\usepackage{times}
\usepackage{subfigure}
\usepackage{mathrsfs}
\usepackage{hyperref}
\usepackage{bbold}
\usepackage{comment}
\usepackage{fancyhdr}
\usepackage{ulem}

\let\oldhat\hat
\renewcommand{\hat}[1]{\oldhat{\mathbf{#1}}}

\newcommand{\beq}{\begin{equation}}
\newcommand{\eeq}{\end{equation}}

\newcommand{\qq}{\mathbf{q}}

\newcommand{\kk}{\mathbf{k}}

\newcommand{\iu}{{i\mkern1mu}}

\newcommand{\BaFeAs}{BaFe$_2$As$_2$}

\newcommand{\gs}{\text{g.s.}}
\newcommand{\zero}{\mathbf{0}}

\begin{document}

\title{Spin Ferroquadrupolar Order in the Nematic Phase of FeSe}
\author{Zhentao Wang}
\author{Wen-Jun Hu}
\author{Andriy H. Nevidomskyy}
\affiliation{Department of Physics and Astronomy, Rice University, Houston, Texas 77005, USA}

\begin{abstract}
We provide evidence that spin ferroquadrupolar (FQ) order is the likely ground state in the nonmagnetic nematic phase of stoichiometric FeSe. By studying the variational mean-field phase diagram of a bilinear-biquadratic Heisenberg model up to the 2nd nearest neighbor, we find the FQ phase in close proximity to the columnar antiferromagnet commonly realized in iron-based superconductors; the stability of FQ phase is further verified by the density matrix renormalization group. The dynamical spin structure factor in the FQ state is calculated with flavor-wave theory, which yields a qualitatively consistent result with inelastic neutron scattering experiments on FeSe at both low and high energies. We verify that FQ can coexist with $C_4$ breaking environments in the mean-field calculation, and further discuss the possibility that quantum fluctuations in FQ act as a source of nematicity.
\end{abstract}

\maketitle

Superconductivity in the iron-based superconductors~\cite{Kamihara2008,Stewart2011} is widely recognized to have spin fluctuations at its origin~\cite{Scalapino2012,Chubukov2012}, as it develops after the suppression of columnar antiferromagnetism (CAFM) by doping or applied pressure on the parent compounds~\cite{LaCruz2008,Rotter2008,Torikachvili2008,Paglione2010}.
The CAFM phase is characterized by the magnetic Bragg peaks at wave vectors $\bm{\mathit Q}_{1,2}=(\pi,0)/(0,\pi)$ in the one-iron Brillouin zone, seen ubiquitously in different families of the iron pnictides and chalcogenides~\cite{LaCruz2008,Lumsden2010,Dai2015}. The discovery of superconductivity in stoichiometric FeSe thus came as a surprise, because the long-range magnetic order is conspicuously absent in this material~\cite{McQueen2009,Bendele2010,Bendele2012,Terashima2015,Watson2015a,Watson2015b}. Another important feature, universally observed across different families of  iron-based superconductors, is the appearance of an electronic nematic phase~\cite{Chu2010, Tanatar2010,Yi2011,Chu2012}, which spontaneously breaks the lattice $C_4$ rotational symmetry. 
Usually, nematicity appears in close proximity to magnetism above the N\'eel temperature; however, in FeSe, the nematic phase appears without any accompanying magnetism and coexists with superconductivity~\cite{Bendele2010,Bendele2012,Terashima2015,Watson2015a}. 
It is thus important to understand the origin of this nonmagnetic nematic phase, in particular, to gain insight into its effect on superconductivity.

It turns out that magnetic order can be induced by applying hydrostatic pressure to FeSe~\cite{Bendele2010,Bendele2012,Terashima2015}. It has also been suggested based on {\it ab initio} calculations that the nonmagnetic phase in FeSe lies in close proximity to the CAFM phase \cite{Subedi2008,Essenberger2012,Heil2014}. Further evidence of proximity to long-range magnetic order comes from inelastic neutron scattering (INS) experiments, which found large spectral weight at wave vectors $\bm{\mathit Q}_{1,2}$~\cite{Rahn2015, Wang2016, QWang2015b, Shamoto2015}. Two natural questions arise: In the theoretical phase diagram, is there a nonmagnetic phase that neighbors on the CAFM? And, furthermore, how does such a nonmagnetic phase give rise to nematicity?

In an attempt to answer these questions, several theoretical scenarios have been proposed for nonmagnetic ground states that may appear as a result of  frustration: a nematic quantum paramagnet~\cite{Wang2015}, a spin quadrupolar state with wave vectors $\bm{\mathit Q}_{1,2}$~\cite{Yu2015}, or a staggered dimer state~\cite{Glasbrenner2015}. In all three cases, the ground state wave function was designed to explicitly break the $C_4$ symmetry, thus resulting in nematicity. 
Alternatively, instead of being the ground state property, nematicity can also be induced as a result of anisotropic thermal~\cite{Fernandes2010,Fernandes2011}, or possibly quantum, fluctuations.

In this Letter, we investigate the frustrated bilinear-biquadratic Heisenberg model used by many authors to model iron pnictides and chalcogenides~\cite{Fang2008,Wysocki2011,Yu2012,Wang2015,Yu2015}, and show that the most likely nonmagnetic state that agrees qualitatively with the INS data on FeSe is the spin ferroquadrupolar (FQ) phase. By using variational mean-field, flavor-wave expansion, and the density matrix renormalization group (DMRG) calculations, we firmly establish that the FQ phase 
is situated in close proximity to the CAFM state in the phase diagram and is readily accessible in the realistic parameter regime of the model. The experimentally observed onset of magnetism in FeSe under applied pressure~\cite{Bendele2010,Bendele2012,Terashima2015} is thus interpreted as the transition between the proposed FQ phase and CAFM.
 The calculated dynamical spin structure factors agree qualitatively with the INS data~\cite{Rahn2015, Wang2016, QWang2015b, Shamoto2015}, exhibiting pronounced maxima of the scattering intensity at the gapped $\bm{\mathit Q}_{1,2}$ points. We note that this is in contrast with the antiferroquadrupolar (AFQ) scenario, which has negligible spectral weight  at these wave vectors~\cite{Yu2015}.
 Furthermore, we demonstrate that FQ order is robust with respect to the $C_4$ symmetry breaking environment, and can thus support nematicity, regardless of its microscopic origin. Additionally, we find that the density-density interactions between $\bm{\mathit Q}_{1,2}$ modes are highly repulsive within the FQ phase and diverge upon approaching the FQ-CAFM phase boundary, providing a scenario in which quantum fluctuations in FQ are the origin of nematicity.

We use a bilinear-biquadratic Heisenberg model~\cite{Fang2008,Wysocki2011,Yu2012,Wang2015,Yu2015} to investigate the ground state properties and spin dynamics:
\begin{equation}\label{eq:model}
\mathcal{H}= \frac{1}{2} \sum_{i,j} J_{ij} \bm{S}_i \cdot \bm{S}_j +  \frac{1}{2}  \sum_{i,j} K_{ij} (\bm{S}_i \cdot \bm{S}_j )^2,
\end{equation}
where $\bm{S}_i$ is the quantum spin-1 operator on site $i$. In the present study, the interactions are limited to the 1st and 2nd nearest neighbors: $J_{ij}=\{J_1, J_2\},\,K_{ij}=\{ K_1, K_2\}$.

The quadrupolar operators are traceless symmetric tensors $Q^{\alpha \beta}\!\equiv\! S^\alpha S^\beta + S^\beta S^\alpha-\tfrac{4}{3} \delta_{\alpha \beta}$ \mbox{($\alpha,\beta\!=\!x,y,z$)}. Only five of these tensors are linearly independent, which are convenient to cast in a 5-vector form: \mbox{$\bm{\mathit Q} \equiv \left( \tfrac{1}{2}(Q^{xx}-Q^{yy}), \tfrac{1}{2\sqrt{3}}(2Q^{zz}-Q^{xx}-Q^{yy} ), Q^{xy}, Q^{yz}, Q^{xz}\right)$}.  The model Eq.~(\ref{eq:model}) can then be rewritten as
\begin{equation}\label{eq:model2}
\mathcal{H}=\frac{1}{2} \sum_{i,j} \left(J_{ij}-\frac{K_{ij}}{2}\right) \bm{S}_i \cdot \bm{S}_j + \frac{1}{4} \sum_{i,j} K_{ij} \left(\bm{\mathit Q}_i \cdot \bm{\mathit Q}_j + \frac{8}{3}\right).
\end{equation}

A time reversal invariant basis for spin-1 is used in this Letter, 
$|\alpha\rangle = \{ \, |x\rangle,\, |y\rangle,\,  |z\rangle \, \}$, defined as a unitary transformation from the regular $|S_z\rangle$ basis:
\begin{equation}
  |x\rangle=\iu \frac{|1\rangle-|\bar{1}\rangle}{\sqrt{2}},\quad
  |y\rangle=\frac{|1\rangle+|\bar{1}\rangle}{\sqrt{2}},\quad
  |z\rangle=-\iu |0\rangle.
  \label{eq:basis}
\end{equation}
An arbitrary single site state can be represented by a unit-length director $\vec{d}_i$, in this basis $| \vec{d}_i\rangle=\sum_\alpha d_i^\alpha |\alpha \rangle $. 

Given a spin state parametrized by director $\vec{d}_i$, the energy of the model Eq.~(\ref{eq:model2}) can be readily calculated at the mean-field level by decoupling $\langle \bm{S}_i \cdot \bm{S}_j  \rangle \approx \langle \bm{S}_i \rangle \cdot \langle \bm{S}_j  \rangle$ 
and similarly for $\langle \bm{\mathit Q}_i \cdot \bm{\mathit Q}_j  \rangle$. 
 Such mean-field decoupling is justified in a minimally entangled long-range order state, for which the wave function can be written in a separable form $|\Psi\rangle = \prod_i |\vec{d}_i\rangle$~\cite{Papanicolaou1988}.
The mean-field ground state energy density is given by
\begin{equation}\label{eq:E0}
E_0 = \frac{1}{2N} \sum_{i,j} \left[ J_{ij} |\langle  \vec{d}_i | \vec{d}_j \rangle |^2 - (J_{ij}-K_{ij}) |\langle  \vec{d}_i | \vec{d}_j^* \rangle |^2 +K_{ij} \right],
\end{equation}
where $N$ stands for the total number of lattice sites.

We then perform a variational search by minimizing Eq.~\eqref{eq:E0} with respect to $\vec{d}_i$, where the directors $\vec{d}_i$ are restricted on $2 \times 2$ and $4 \times 4$ unit cells with periodic boundary condition. The purely quadrupolar states are identified with vanishing magnetic moment: $\langle \bm{S}_i \rangle \equiv 2\, \text{Re}[ \vec{d}_i ] \times \text{Im}[ \vec{d}_i ]=0, \forall i$. Among the quadrupolar states, one distinguishes a FQ phase, with all directors parallel, and more general AFQ phases with noncollinear directors.
The familiar  magnetic phases correspond to dipolar moment $| \langle \bm{S}_i \rangle |=1, \forall i$ with a spin structure factor characterized by the Bragg peaks.
In general, one also encounters states that contain a mixture of magnetic and quadrupolar moments with $0 < |\langle \bm{S}_i \rangle |<1$ on all sites, or states that have purely magnetic or quadrupolar moments only on partial sites, or even so-called semiordered states with undetermined $ |\langle \bm{S}_i \rangle |$ ~\cite{Papanicolaou1988}.

\begin{figure}[!b]
\includegraphics[width=0.45\textwidth]{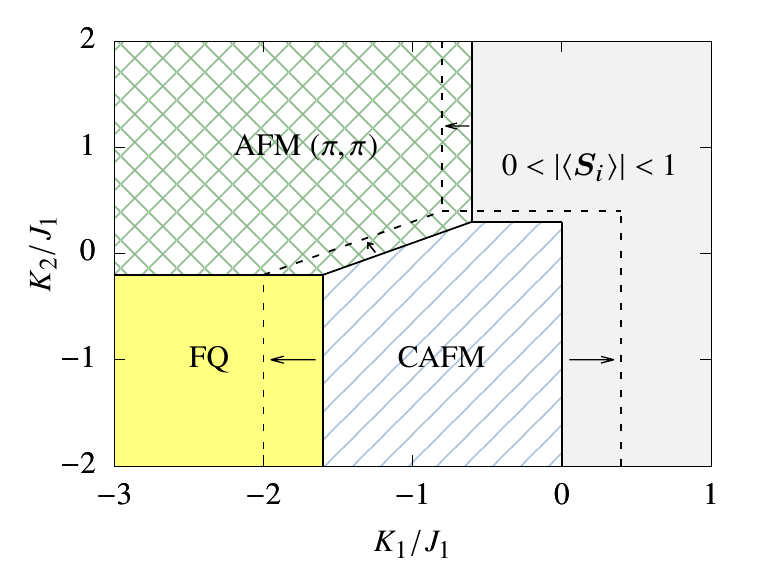}
\caption{Variational mean-field phase diagram of the Hamiltonian Eq.~\eqref{eq:model} with $J_1=1,J_2=0.8$, and periodic boundary condition ($2 \times 2$ and $4 \times 4$ unit cells yield exactly the same results)~\cite{mis_ed}. The dashed lines denote shifted phase boundaries when breaking $C_4$ symmetry in Eq.~\eqref{eq:model} by hand, using $J_1^{x,y} = (1\pm 0.2)J_1$.
}\label{Fig1}
\end{figure} 

The  variational mean-field phase diagram is given in Fig.~\ref{Fig1}, obtained for antiferromagnetic $J_1>0$ and $J_2/J_1=0.8$, which were deduced by fitting the INS spectra for \BaFeAs~\cite{Harriger2011} to the $J_1-J_2-K_1$ spin model~\cite{Wysocki2011,Yu2012}. 
Because of the fact that FeSe lies in proximity to CAFM, we do not expect its parameters to deviate dramatically from those deduced in Refs.~\cite{Wysocki2011,Yu2012}, and we have also verified that the magnetic and quadrupolar phases in Fig.~\ref{Fig1} are robust to small variations of $J_2/J_1$.
Remarkably, Fig.~\ref{Fig1} shows that the only nonmagnetic phase in close proximity to CAFM is the FQ phase, with both phases realized at negative biquadratic interaction $K_1$. We note  that $K_1<0$ is generically expected from the fitting of the INS spectra in the iron pnictides/chalcogenides~\cite{Wysocki2011,Yu2012}, with the ratio $|K_1|/J_1$ of order 1, consistent with the location of CAFM region in Fig.~\ref{Fig1}. The large negative $K_1$ is also expected from the spin crossover model by Chaloupka and Khaliullin~\cite{Chaloupka2013}, which also incorporates the FQ and CAFM phases; and large $|K_1|$ also naturally arises within the Kugel-Khomskii type models when the orbitals order inside the nematic phase~\cite{Ergueta2015}.
No other purely quadrupolar phases were found; in particular, the AFQ($\pi,0$)/($0,\pi$) phase, expected to be realized for positive $K_2$~\cite{Yu2015} turns out to be unstable to the admixture of the magnetic moment, resulting in a mixed magnetic or quadrupolar state with $0 < |\langle \bm{S}_i \rangle |<1$ (gray region in Fig.~\ref{Fig1})~\cite{supp}.

\begin{figure}[!b]
\includegraphics[width=0.48\textwidth]{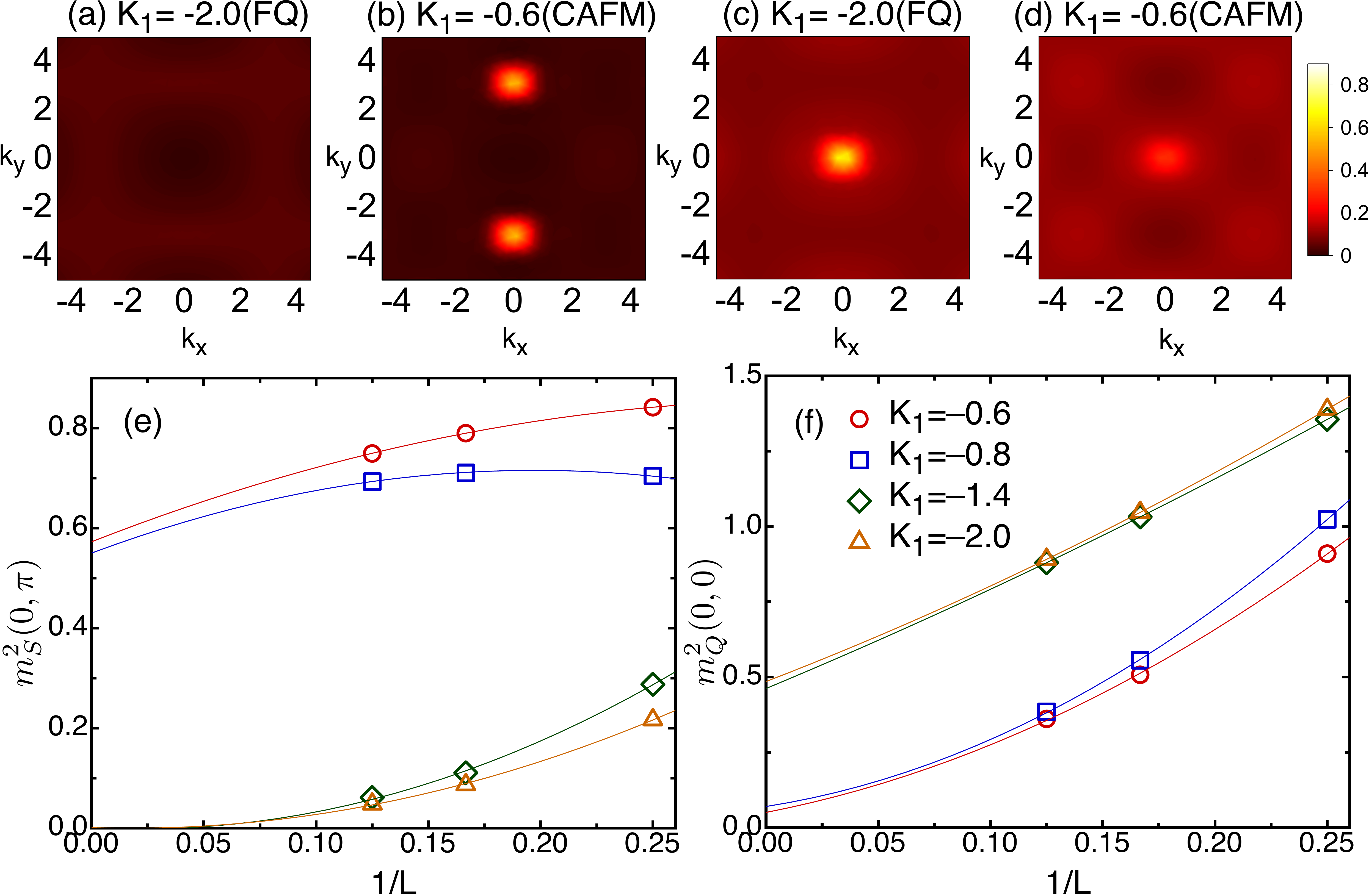}
\caption{\label{Fig2}Static spin and quadrupolar structure factors obtained from DMRG on RC$L\!\! -\!\! 2L$ cylinders with $J_1=1, J_2=0.8, K_2=-1$. 
(a),(b) $m_S^2(\bm{q})$ for $L=8$.
(c),(d) $m_Q^2(\bm{q})$ for $L=8$.
(e),(f) Finite-size scaling of $m^{2}_{S}[\bm{q}=(0,\pi)]$ and $m^{2}_{Q}[\bm{q}=(0,0)]$ as a function of the inverse cylinder width, where the lines are guides to the eye.}
\end{figure}

Since the variational mean-field calculation only takes into account minimally entangled mean-field states, 
the results in Fig.~\ref{Fig1} may be energetically unfavorable upon quantum fluctuations. 
To verify the stability of the FQ phase, we have performed the SU(2) DMRG calculations~\cite{White1992,mcculloch2002,gong2014square,gong2014kagome} on $L\!\! \times\!\! 2L$ rectangular cylinders with $L=(4,6,8)$~\footnote{$L$ represents the size of y direction which has periodic boundary condition} near the mean-field FQ-CAFM phase boundary.
We keep up to $6000$ SU$(2)$ states, leading to truncation errors less than $10^{-5}$ in all data points presented in this Letter. 
In Fig.~\ref{Fig2}, we show both the static spin and quadrupolar structure factors, 
defined as $m^{2}_{S}(\bm{q})=\frac{1}{L^4}\sum_{ij} \langle {\bf S}_{i}\cdot {\bf S}_{j}\rangle e^{i\bm{q}\cdot(\bm{r}_i-\bm{r}_j)}$
and $m^{2}_{Q}(\bm{q})=\frac{1}{L^4}\sum_{ij} \langle \bm{\mathit Q}_{i}\cdot \bm{\mathit Q}_{j}\rangle e^{i\bm{q}\cdot(\bm{r}_i-\bm{r}_j)}$
(where $i,j$ are only partially summed on $L \times L$ sites in the middle of the cylinder, in order to reduce boundary effects~\cite{white2007,Yan2011,gong2014square,Gong2015_honeycomb}). 
Figures.~\ref{Fig2}(a) and \ref{Fig2}(b) show the results for $m_S^2(\bm{q})$ in the FQ and CAFM phases, respectively; Figs.~\ref{Fig2}(c) and \ref{Fig2}(d) depict $m_Q^2(\bm{q})$ in these two phases.
Since $m_S^2(\bm{q})$ and $m_Q^2(\bm{q})$ are maximized near $(0,\pi)$ and $(0,0)$, respectively, we fix $\bm{q}$ at these two momenta, and perform finite size scaling analysis of $m^{2}_{S}(\bm{q})$ and $m^{2}_{Q}(\bm{q})$ in Figs.~\ref{Fig2}(e) and \ref{Fig2}(f). 
For large negative $K_1$, it is clearly shown that the $m_S^2(0,\pi)$ is suppressed from $L=4$ to $8$, and vanishes in the thermodynamic limit by extrapolation; while $m_Q^2(0,0)$ remains finite, confirming FQ as the underlying phase. 
For small negative $K_1$, $m_S^2(0,\pi)$ remains finite in the thermodynamic limit, confirming the corresponding phase to be CAFM.
We note that the DMRG yields a larger FQ region with the FQ-CAFM boundary found at $K_1 > -1.4$, compared  to the mean-field prediction of $K_1^c=-1.6$ in Fig.~\ref{Fig1}.

\begin{figure}[!b]
\includegraphics[width=0.45\textwidth]{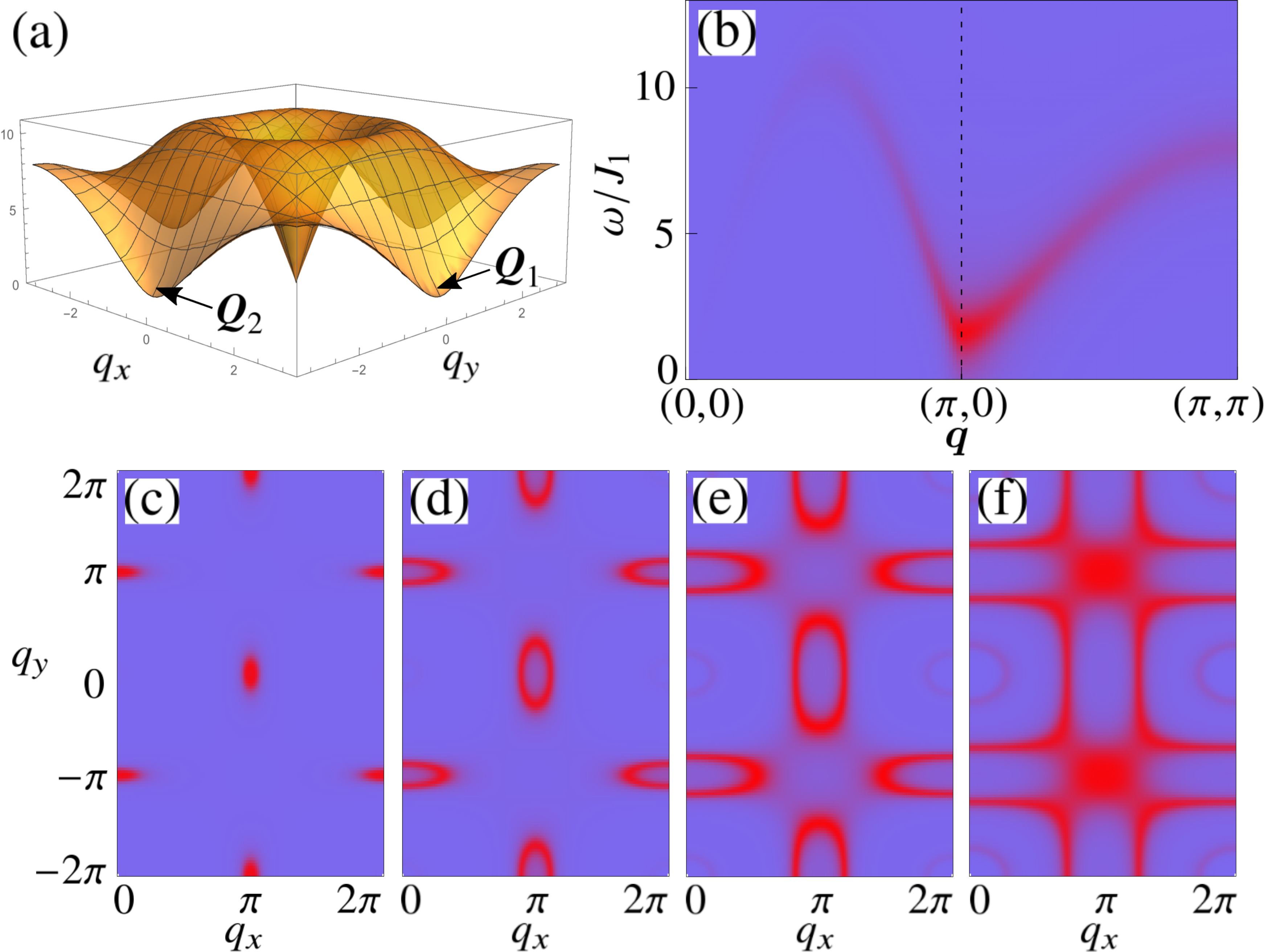}
\caption{\label{Fig3}Dispersion and dynamical spin structure factor in the FQ phase obtained from flavor-wave calculation with $J_1=1, J_2=0.8, K_1= -1.65,K_2=-0.8$. (a) Dispersion plotted in the 1st BZ. (b) Energy-momentum dependence of $S(\qq,\omega)$. (c)--(f) Constant-energy cuts of $S(\qq,\omega)$ in $\bm{q}$ space. (c) $\omega/J_1=2$. (d) $\omega/J_1=4$. (e) $\omega/J_1=6$. (f) $\omega/J_1=8$. A Lorentzian broadening factor $\lambda=0.8J_1$ is used for approximating the delta functions.}
\end{figure}

Having established FQ as a stable nonmagnetic phase in close proximity to CAFM, we turn to the analysis of its magnetic exictations. We use the flavor-wave technique, which represents the local spin and quadrupolar operators $\mathscr{O}_i$ in terms of three flavors of Schwinger bosons in the fundamental representation of SU$(3)$~\cite{Papanicolaou1988, Tsunetsugu2006, Lauchli2006, Muniz2014}: $\mathscr{O}_i =\sum_{\alpha \beta} b_{i, \alpha}^\dagger O_i^{\alpha \beta} b_{i, \beta}$, subject to the constraint $\sum_\alpha b_{i, \alpha}^\dagger b_{i, \alpha}=1$. 
The quadrupolar solution corresponds to the Bose-Einstein condensation of the appropriate boson (labeled $b_z$), and the remaining two flavors capture  both spin and quadrupolar excitations~\cite{Papanicolaou1988, Tsunetsugu2006, Lauchli2006}. Expanding $b_{i,z}^\dagger = b_{i,z}=\sqrt{1-b_{i,x}^\dagger b_{i,x}-b_{i,y}^\dagger b_{i,y}}$ and keeping up to bilinear terms in the Hamiltonian Eq.~\eqref{eq:model2}, it can be diagonalized by the standard Bogoliubov transformation $\alpha_{\bm{q},a}=\cosh \theta_{\bm{q}} b_{\bm{q},a}-\sinh \theta_{\bm{q}} b_{-\bm{q},a}^\dagger$, yielding (up to a  constant)~\cite{supp}
\begin{equation}\label{Eq:diag}
\mathcal{H}_{\text{fw}}=\sum_{a=x,y} \sum_{\bm{q}} \omega_{\bm{q},a} (\alpha_{\bm{q},a}^\dagger \alpha_{\bm{q},a}+1/2),
\end{equation}
where dispersion $\omega_{\bm{q},a}$ are degenerate in flavor index $a=\{x,y\}$, shown in Fig.~\ref{Fig3}(a). Since the FQ phase spontaneously breaks the spin-rotational symmetry, there are two gapless Goldstone modes at $\qq\!=\!\zero$.  However, there is no Bragg peak as the dynamical spin structure factor $S(\qq,\omega)$ shown in Fig.~\ref{Fig3}(b) has a vanishing spectral weight ($\propto |\bm{q}|$) at $\qq\!=\!0,\omega\!=\!0$  because of the conservation of time reversal symmetry in quadrupolar states~\cite{Lauchli2006,Tsunetsugu2006,Smerald2013,Smerald2015}.
In Fig.~\ref{Fig3}(b), we see large spectral weight at $\bm{\mathit Q}_{1,2}$ at low energy due to the proximity to the CAFM phase. The spectral weight further shifts towards $(\pi,\pi)$ when increasing $\omega$ [see Figs.~\ref{Fig3}(c)-\ref{Fig3}(f)], closely tracking the INS results on FeSe~\cite{Rahn2015,Wang2016, QWang2015b, Shamoto2015}. We  note that in the AFQ $(\pi,0)/(0,\pi)$ phase proposed in Ref.~\onlinecite{Yu2015}, 
one would expect Goldstone modes with zero spectral weight at $\bm{\mathit Q}_{1,2}$, which would contradict the large-intensity dispersing feature  near $\bm{\mathit Q}_{1,2}$ found in the INS data on FeSe.

Having demonstrated that the FQ phase is indeed consistent with the INS results on FeSe~\cite{Rahn2015,Wang2016, QWang2015b, Shamoto2015}, we now ask further whether the FQ phase can coexist with nematicity observed in FeSe.
We apply $C_4$ breaking exchange anisotropy in Eq.~\eqref{eq:model}, using $J_1^{x,y} = (1\pm 0.2)J_1$ in the variational mean-field calculation. This results in the shift of the phase boundaries (shown with dashed lines in Fig.~\ref{Fig1}) and, although the FQ phase shrinks slightly, it clearly remains stable in a large portion of the mean-field phase diagram.

We now turn to the microscopic origin of nematicity in FeSe--can FQ order be the reason for the discrete $C_4$ symmetry breaking?
Unlike other proposals starting with nematic spin wave functions in the ground state~\cite{Wang2015,Yu2015,Glasbrenner2015}, in the flavor wave theory up to bilinear terms in Eq.~(\ref{Eq:diag}), the spin correlations in the FQ phase are $C_4$ symmetric.
This does not mean that the FQ ground state cannot spontaneously break this symmetry and, in fact, it turns out that higher order interactions (mode-mode coupling) become increasingly important when approaching the FQ-CAFM phase boundary. 
Collecting up to the 4th order terms in the flavor wave theory~\cite{supp}, we obtain $\mathcal{H}_{\text{4th}} = \mathcal{H}_{\text{fw}} + \mathcal{H}_{\text{int}}$ with
\begin{equation}\label{Eq:quartic}
\mathcal{H}_{\text{int}} \!=\!\! \frac{1}{N} \sum_{abcd} \sum_{\kk_1,\kk_2,\qq} \!\! V_{ab}^{cd}(\kk_1,\kk_2,\qq) \alpha^\dag_{\kk_1+\qq,a} \alpha^\dag_{\kk_2-\qq,b} \alpha_{\kk_2,c} \alpha_{\kk_1,d},
\end{equation}
where only five combinations of $\{abcd\}$ are nonzero: $\{$xxxx$\},\{$yyyy$\},\{$xxyy$\},\{$yyxx$\}$, and $\{$xyyx$\}$. Above, only particle number conserving terms have been kept for simplicity.

In terms of Schwinger bosons, we can define a nematic order parameter as $\langle \Delta \rangle = \sum_a \langle n_{\bm{\mathit Q}_1,a}-n_{\bm{\mathit Q}_2,a} \rangle$, where $\langle \ldots \rangle$ denotes the expectation value in the full interacting Hamiltonian $\mathcal{H}_{\text{4th}}$, and $n_{\bm{q},a}=\alpha_{\bm{q},a}^\dagger \alpha_{\bm{q},a}$ is the boson density operator of flavor $a$ at momentum $\bm{q}$. 
If we stop at the quadratic level of flavor wave theory, then $\langle \Delta \rangle_{\text{fw}} \equiv 0$ due to the Bose-Einstein condensation at $\bm{q}=(0,0)$. Once interactions are taken into account in $H_{\text{4th}}$, the condensate will become depleted, resulting in a finite boson density at the local minima $\bm{\mathit Q}_{1,2}$  of the spectrum in Fig.~\ref{Fig3}(a) and thus making it possible, in principle,  that $\langle \Delta \rangle \neq 0$. To see how this may occur, we 
consider the density-density interactions between the $\bm{\mathit Q}_{1,2}$ modes, which can be extracted from Eq.~\eqref{Eq:quartic} as
\begin{equation}
\mathcal{H}_{\text{int}} = \tilde{V} (n_{\bm{\mathit Q}_1,x} n_{\bm{\mathit Q}_2,x} + n_{\bm{\mathit Q}_1,y} n_{\bm{\mathit Q}_2,y}) 
+ \tilde{V}^\prime  n_{\bm{\mathit Q}_1,x} n_{\bm{\mathit Q}_2,y} + \ldots,
\end{equation}
where the intraflavor and interflavor interactions $\tilde{V}$  and $\tilde{V}^\prime$  are expressed~\cite{supp} through  $V_{ab}^{cd}(\kk_1,\kk_2,\qq)$ in Eq.~(\ref{Eq:quartic}).

\begin{figure}[!tb]
\includegraphics[width=0.45\textwidth]{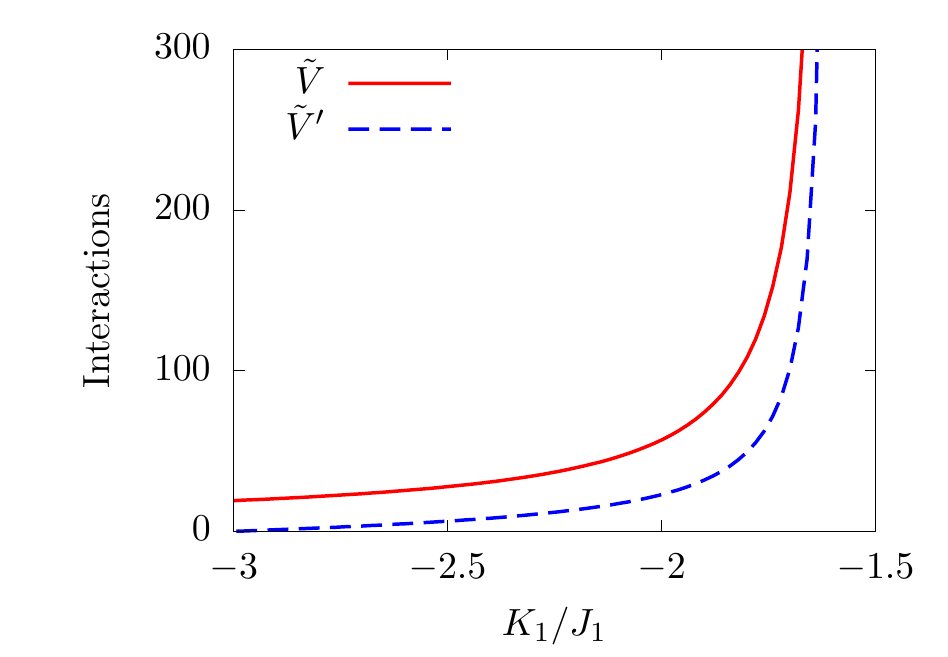}
\caption{\label{Fig4}The density-density interactions between the $\bm{\mathit Q}_{1,2}$ modes when approaching the FQ-CAFM phase boundary $K_1^c=-1.6$.
The parameters used in this plot are $J_1=1,J_2=0.8$, and $K_2=-0.8$.}
\end{figure}

The values of $\tilde{V}$ and $\tilde{V}^\prime$ are plotted in Fig.~\ref{Fig4}. Intriguingly, they are repulsive in the region $K_1>-3$, and diverge when approaching the FQ-CAFM phase boundary at $K_1^c=-1.6$, resulting in a $C_4$ symmetry-breaking imbalance in boson occupation $n_{\bm{\mathit Q}_1}\neq n_{\bm{\mathit Q}_2}$.
Since sufficiently strong (not necessarily diverging) interactions can commonly trigger diverging susceptibilities,
we expect the renormalized nematic susceptibility to diverge before reaching the FQ-CAFM phase boundary, resulting in a finite nematic window  $K_1^N<K_1< K_1^c$ inside the FQ phase. The existence of such a window should be carefully verified by further analytical and numerical efforts, which will be a subject of future work. We note that while the present study is limited to second-neighbor interactions, our mean-field analysis shows that  inclusion of third neighbor $K_3(\mathbf{S}_i\cdot \mathbf{S}_j)^2$ term with  $K_3<0$ will further favor FQ over magnetic phases~\cite{supp}, possibly leading to a wider nematic region.

Direct experimental measurements of quadrupolar orders are typically difficult, due to the negligible spectral weight of the spin structure factor near the ordering wave vector. A possible way to visualize such ``ghost" modes is by applying a magnetic field: the degeneracy of the two flavors will be lifted, and one of the Goldstone modes acquires a gap and 
a visible spectral weight~\cite{Smerald2013,Smerald2015}, as we demonstrate in Ref.~\cite{supp}. 
The quadrupolar orders can also be measured by  Raman scattering, which is able to couple to spin and quadrupolar operators by tuning light polarization and incoming light frequency, thus showing qualitatively different features for magnetic and quadrupolar phases~\cite{Michaud2011}.
More direct evidence can be gained from the quadrupolar  structure factor, which should exhibit Bragg peaks at the ordering wave vector~\cite{Smerald2013}, and, in principle, can be measured by resonant inelastic x-ray scattering experiments~\cite{Kuramoto2009, Ament2011}.  

In the present work, the effect of conduction electrons on the spin dynamics has been neglected for simplicity sake; the latter lead to an additional broadening of the INS features due to the Landau damping~\cite{Yu2012}, but do not otherwise impact our conclusions.


In summary, we showed that the FQ phase lies in close proximity to CAFM in the phase diagram of a bilinear biquadratic spin-1 model and that it is stable in a realistic range of the model parameters, as verified by both the mean-field and DMRG methods. 
 The dynamical spin structure factor $S(\bm{q},\omega)$ inside the FQ phase is shown to be qualitatively consistent with the recent INS results on FeSe. While at the quadratic level the FQ ground state does not explicitly break the $C_4$ lattice symmetry, we demonstrate that the quantum fluctuations result in repulsive density-density interactions between $\bm{\mathit Q}_{1,2}$ magnon modes, whose strength diverges on approaching the FQ-CAFM phase boundary. 
This suggests the existence of a finite window inside the nonmagnetic FQ phase where the C$_4$ symmetry is spontaneously broken.
Further studies are necessary to establish such a nematic window unequivocally; however, even if the nematicity is driven by other sources (for example, local strains due to lattice imperfections, or orbital ordering, as proposed in the light of recent nuclear magnetic resonance~\cite{Bohmer2014, Baek2014} and ARPES~\cite{Shimojima2014} experiments), the incipient nematic order will couple to the symmetry-breaking quantum fluctuations that we found in the FQ phase. Our calculations show that the FQ order is robust with respect to such $C_4$ breaking environments and can coexist with nematicity.

We would like to thank Shou-Shu~Gong for helpful discussions and providing the DMRG code. 
A.~H.~N. thanks Karlo Penc and Nic Shannon for fruitful discussions and the hospitality of the Institute for Solid State Physics (ISSP) at the University of Tokyo, where part of this work was performed.
A.~H.~N. is also grateful for the hospitality of the Aspen Center for Physics which is supported by NSF Grant No.~PHY-1066293.
Computational resources were provided by the Big-Data Private-Cloud Research Cyberinfrastructure MRI-award funded by NSF under Grant No.~CNS-1338099 and by Rice University;
and by the Extreme Science and Engineering Discovery Environment (XSEDE)~\cite{Towns2014}, which is supported by NSF Grant No.~ACI-1053575.
Z.~W. was supported by the Welch Foundation Grant No.~\mbox{C-1818}. A.~H.~N. and W.~H. acknowledge the support of the NSF CAREER Grant No.~DMR-1350237. 
W.~H. also acknowledges the support of NSF Grant No.~DMR-1309531. 

{\it Note added in Proof.--} 

Recently, we became aware of a study on the $J_{ij}-K_{ij}$ model up to the third neighbors~\cite{Lai2016}. In addition to the FQ state, those authors also find evidence of the AFQ $(\pi,0)/(0,\pi)$ phase stabilized by a large negative $K_3$, which sits far away from the CAFM phase in the theoretical phase diagram.

{\it Erratum.--} 

When comparing theoretical predictions of the antiferroquadrupolar (AFQ) order proposed in Ref.~\cite{Yu2015} with the inelastic neutron scattering (INS) experiment, we remark that the AFQ order has dispersing Goldstone modes emanating from $\bm{\mathit Q}_{1,2}=(\pi,0)/(0,\pi)$ wave-vectors. Since the INS spectral function always contains instrumental and damping broadening in both momentum and energy, it is possible to interpret the INS data in Ref.~\cite{QWang2015b} as consistent with the AFQ scenario, as stressed by the authors of Ref.~\cite{Lai2016}. Additionally, the $\bm{q}$-integrated low-energy spectral weight near the wavevector $(\pi,0)$ (the so-called ``stripe" component in Fig.~4b of Ref.~\cite{QWang2015b}), appears linear in energy from the threshold energy just above the range under the influence of superconductivity (about 10 meV) up to about 40 meV. Such a linear dependence is consistent with the expected behavior in the AFQ phase, because this phase preserves the time-reversal symmetry and the low energy spin spectral weight of the quadrupolar Goldstone mode scales linearly with energy. Such a linear dependence is also expected in the FQ scenario when the spectral weight is integrated over momentum near the wavevector $(0,0)$, which remains to be tested experimentally.

\setcounter{figure}{0}
\renewcommand{\thefigure}{S\arabic{figure}}
\setcounter{equation}{0}
\renewcommand{\theequation}{S\arabic{equation}}

\begin{center}
  {\bf ---Supplemental Material---}
\end{center}

\section{List of Mean-field energies}
The mean-field ground state energy density $E_0$ is given by Eq.~(4) in the main text. For some purely dipolar and purely quadrupolar phases of interest, their $E_0$ can be written down explicitly up to the 3rd nearest neighbor:
\begin{eqnarray}\label{Eq:biased}
\text{FM}: E_0 &=& 2J_1 + 2J_2  + 2J_3 + 2K_1 + 2K_2 + 2K_3; \nonumber\\
\text{AFM} (\pi,0)/(0,\pi): E_0 &=& -2J_2  + 2J_3 + 3K_1 + 4K_2 + 2K_3; \nonumber \\
\text{AFM} (\pi,\pi): E_0 &=& -2J_1 + 2J_2 + 2J_3 + 4K_1 +2K_2 + 2K_3 ; \nonumber \\
\text{FQ}: E_0 &=& 4K_1 + 4K_2 + 4K_3; \label{eq:energies}\\
\text{AFQ} (\pi,0)/(0,\pi): E_0 &=& 3K_1 + 2K_2 + 4K_3; \nonumber \\
\text{AFQ} (\pi,\pi): E_0 &=& 2K_1 + 4K_2 + 4K_3. \nonumber
\end{eqnarray}

\begin{figure}[!ht]
\includegraphics[width=0.4\textwidth]{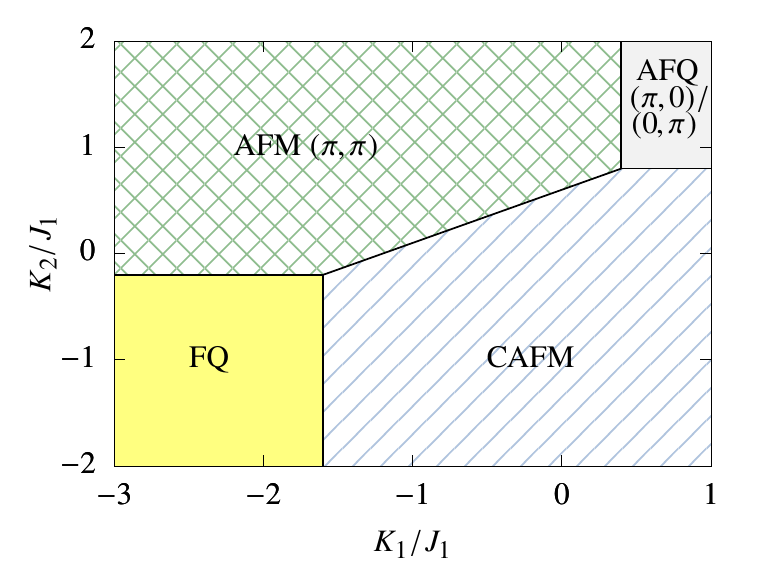}
\caption{\label{FigS1}
 Biased mean-field phase diagram obtained from comparing energies listed in Eq.~\eqref{Eq:biased}, with $J_1=1,J_2=0.8$, and $J_3=K_3=0$.
}
\end{figure}

By restricting ourselves to consider only theses phases listed in Eq.~\eqref{eq:energies}, we can obtain a {\it biased} mean-field phase diagram Fig.~\ref{FigS1}. We note that the AFQ $(\pi,0)/(0,\pi)$ phase in Fig.~\ref{FigS1} is energetically unfavorable in the full variational treatment, and will be replaced by the region $0<|\bm{S}_i|<1$ in Fig.~1 in the main text.

\section{Flavor wave calculation in FQ}
The flavor wave theory is commonly used for spin-1 Hamiltonian, and can be found for example in Ref.~\onlinecite{Papanicolaou1988, Lauchli2006, Tsunetsugu2006}. In this section we put down the details of the flavor wave calcualtion for the model Eq.~(1) defined in the main text. 

Consider Hamiltonian Eq.~(1) with an applied magnetic field $-h \sum_i S_i^z$, then the directors in the FQ phase become:
\begin{equation}\label{Eq:director}
\vec{d}_i = \left( \cos (\mu/2),i \sin (\mu/2), 0 \right),
\end{equation}
where $\mu$ is determined by minimizing Eq.~(4) in the main text, which in an FQ state gives:
\begin{equation}
\sin \mu = \frac{h}{4(J_1+J_2-K_1-K_2)}.
\end{equation}

The local operators are represented by the SU$(3)$ Schwinger bosons in the fundamental representation, described in the main text. Further, we perform a unitary transformation according to the directors in the magnetic field:
\begin{subequations}
\begin{align}
\tilde{\bm{b}}_{i} &= \mathcal{V}_i^\dagger \bm{b}_i, \\
\tilde{S}_i^\nu &= \mathcal{V}_i^\dagger S_i^\nu \mathcal{V}_i, \\
\tilde{Q}_i^\nu &= \mathcal{V}_i^\dagger Q_i^\nu \mathcal{V}_i,
\end{align}
\end{subequations}
where the transformation matrix $\mathcal{V}_i$ is identical on all sites $i$ in an FQ state:
\begin{equation}
\mathcal{V}_i =
\begin{pmatrix}
i \sin (\mu/2) & 0 & \cos (\mu/2) \\
\cos (\mu/2) & 0 & i \sin (\mu/2) \\
0 & 1 & 0
\end{pmatrix}.
\end{equation}

The third component of $\bm{b}_i$ is condensed, by expanding $\tilde{b}_{i,3}^\dagger \!=\! \tilde{b}_{i,3} \! =\! \sqrt{1-\tilde{b}_{i,1}^\dagger \tilde{b}_{i,1}-\tilde{b}_{i,2}^\dagger \tilde{b}_{i,2}} $. The Hamiltonian expanded up to quadratic level can be diagonalized by the Bogoliubov transformation  $\alpha_{\bm{q},a}=\cosh \theta_{\bm{q},a} b_{\bm{q},a}-\sinh \theta_{\bm{q},a} b_{-\bm{q},a}^\dagger$, up to a constant gives:
\begin{equation}
\mathcal{H}_{\text{fw}}=\sum_{a=1,2} \sum_{\bm{q}} \omega_{\bm{q},a} (\alpha_{\bm{q},a}^\dagger \alpha_{\bm{q},a}+1/2),
\end{equation}
where the dispersion $\omega_{\bm{q},a}$:
\begin{equation}
\omega_{\bm{q},a} =\sqrt{(2t_{aa}(\bm{q})+\lambda_{aa})^2-4\Delta_{aa}(\bm{q})^2},
\end{equation}
and the Bogoliubov coefficents:
\begin{equation}
\tanh 2 \theta_{\bm{q},a} = -\frac{2 \Delta_{aa}(\bm{q})}{2 t_{aa}(\bm{k})+\lambda_{aa}},
\end{equation}
where $t_{ab}(\bm{q})$, $\Delta_{ab}(\bm{q})$ and $\lambda_{ab}$ are $2 \times 2$ diagonal matrices in FQ:
\begin{subequations}
\begin{align}
t_{11}(\bm{q}) &= (J_1 \cos^2 \mu + K_1 \sin^2 \mu) (\cos q_x + \cos q_y) \nonumber \\
&\quad + 2(J_2 \cos^2 \mu +K_2 \sin^2 \mu) \cos q_x \cos q_y, \\
t_{22}(\bm{q}) &= J_1 (\cos q_x + \cos q_y) + 2J_2 \cos q_x \cos q_y, \\
\Delta_{11}(\bm{q}) &= (K_1-J_1) \cos^2 \mu \,(\cos q_x + \cos q_y) \nonumber \\
&\quad +2 (K_2-J_2) \cos^2 \mu \, \cos q_x \cos q_y, \\
\Delta_{22}(\bm{q}) &= (K_1-J_1) \cos \mu \,(\cos q_x + \cos q_y) \nonumber \\
&\quad +2 (K_2-J_2) \cos \mu \, \cos q_x \cos q_y, \\
\lambda_{11}&=-8(J_1+J_2-K_1-K_2) \sin^2 \mu \nonumber \\
&\quad  -4(K_1+K_2) + 2h \sin \mu, \\
\lambda_{22} &=-4(J_1+J_2-K_1-K_2) \sin^2 \mu \nonumber \\
&\quad -4(K_1+K_2) + h \sin \mu.
\end{align}
\end{subequations}

The dynamical spin structure factor at $T=0$ is defined as:
\begin{equation} \label{Eq:sqw}
S^{\alpha\beta}(\qq,\omega) \!\! =\!\! \sum_f \langle \gs|S^\alpha(\qq)|f\rangle \langle f|S^\beta(-\qq)|\gs\rangle \delta(\omega-E_f+E_g).
\end{equation}

\begin{figure}[!tb]
\includegraphics[width=0.4\textwidth]{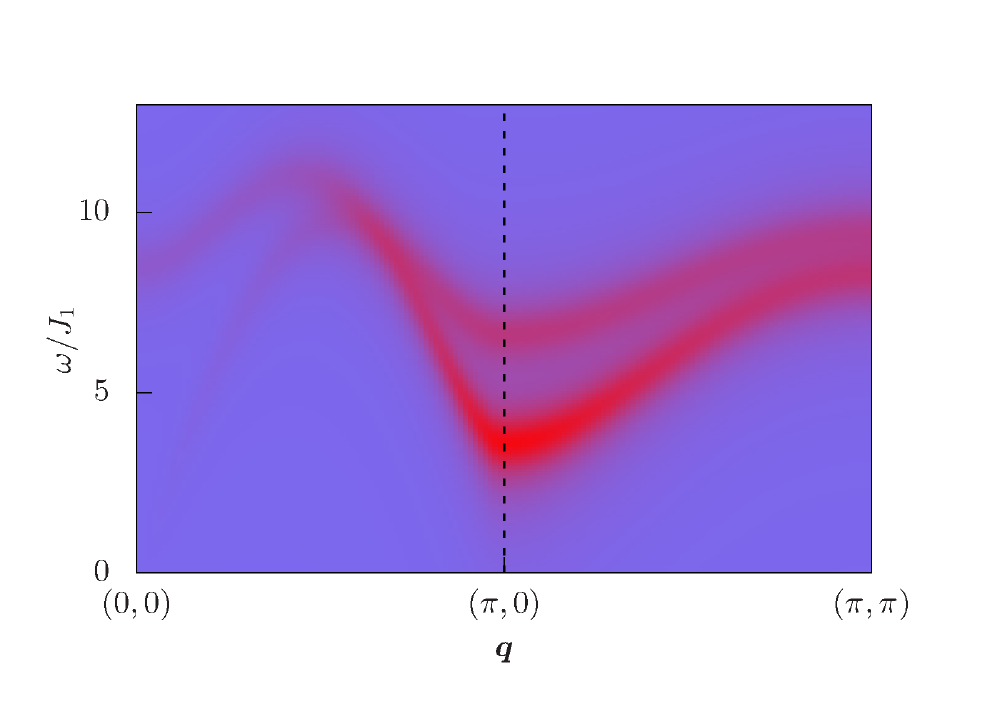}
\caption{\label{FigS2}
Dynamical spin structure factor in the FQ phase with finite magnetic field, plotted using $J_1=1, J_2=0.8, K_1=-1.65, K_2=-0.8, h=2(J_1+J_2-K_1-K_2)$, and a Lorentzian broadening factor $\lambda=0.8J_1$.
}
\end{figure}

The spin operators in Eq.~\eqref{Eq:sqw} are represented by the SU$(3)$ bosons, keeping only the linear order terms (neglecting the two-magnon continum and the constant background):
\begin{subequations}
\begin{align}
S^x(\bm{q}) &= -\sin \frac{\mu}{2} \left( \tilde{b}_{-\bm{q},2}^\dagger + \tilde{b}_{\bm{q},2} \right), \\
S^y(\bm{q}) &= -i \cos \frac{\mu}{2} \left( \tilde{b}_{-\bm{q},2}^\dagger -\tilde{b}_{\bm{q},2} \right), \\
S^z(\bm{q}) &= i \cos \mu \, \left( \tilde{b}_{-\bm{q},1}^\dagger -\tilde{b}_{\bm{q},1}   \right).
\end{align}
\end{subequations}

Then Eq.~\eqref{Eq:sqw} can be written down explictily:
\begin{subequations}
\begin{align}
S^{xx}(\bm{q},\omega)\! &=\! \sin^2 \frac{\mu}{2}  \frac{2t_{22}(\bm{q})\! + \! \lambda_{22}\! - \! 2\Delta_{22}(\bm{q})}{\omega_{\bm{q},2}} \delta(\omega \! -\! \omega_{\bm{q},2}), \\
S^{yy}(\bm{q},\omega) \! &=\! \cos^2 \frac{\mu}{2}  \frac{2t_{22}(\bm{q})\! +\! \lambda_{22}\! + \! 2\Delta_{22}(\bm{q})}{\omega_{\bm{q},2}} \delta(\omega \! -\! \omega_{\bm{q},2}), \\
S^{zz}(\bm{q},\omega) \! &=\! \cos^2 \frac{\mu}{2} \frac{2t_{11}(\bm{q})\! +\! \lambda_{11}\! + \! 2\Delta_{11}(\bm{q})}{\omega_{\bm{q},1}} \delta(\omega\! -\! \omega_{\bm{q},1}).
\end{align}
\end{subequations}

The resulting dipolar spin structure factor at zero field is shown in Fig.~3 in the main text. At finite field, the degeneracy between the two Goldstone modes splits, shown in Fig.~\ref{FigS2}.

\section{Quartic Interactions}

In Eq.~(7) in the main text, the density-density interactions between $\bm{\mathit Q}_{1,2}$ are extracted from Eq.~(6) as:
\begin{subequations}
\begin{align}
\tilde{V} &=V_{xx}^{xx}(\bm{\mathit Q}_1,\bm{\mathit Q}_2,(0,0))+V_{xx}^{xx}(\bm{\mathit Q}_1,\bm{\mathit Q}_2,(\pi,\pi)), \\
\tilde{V}^\prime &=V_{xy}^{yx}(\bm{\mathit Q}_1,\bm{\mathit Q}_2,(0,0)),
\end{align}
\end{subequations}
where the intra-flavor interactions are identical for $x$ and $y$ flavors.

\begin{figure}[!htb]
\includegraphics[width=0.4\textwidth]{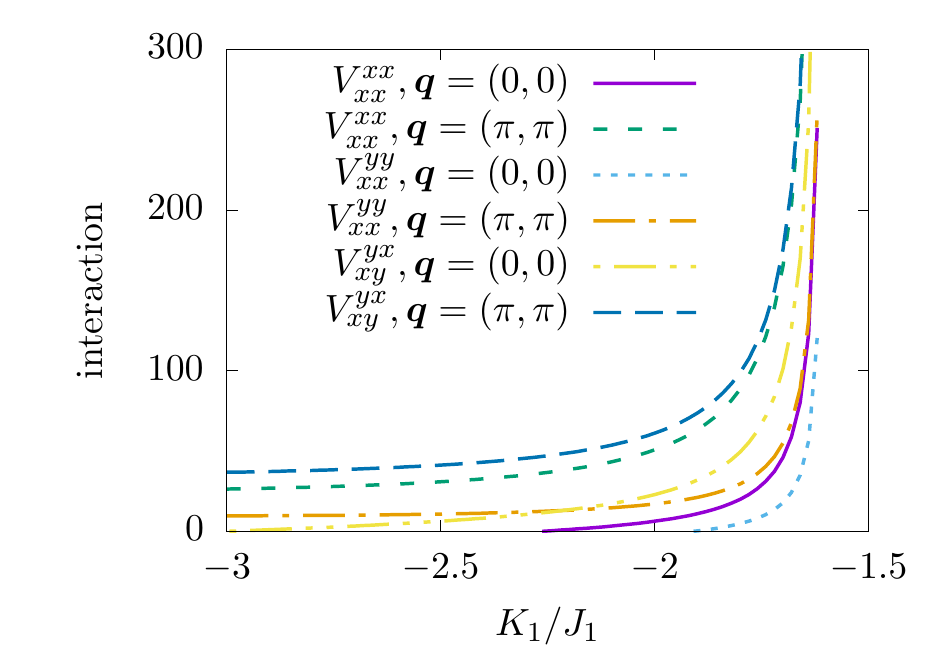}
\caption{\label{FigS3}
Quartic interactions by fixing incoming momenta at $\bm{\mathit Q}_1,\bm{\mathit Q}_2$, and exchange momenum $\bm{q}=(0,0)$ or $\bm{q}=(\pi,\pi)$. Plotted using $J_1=1, J_2=0.8, K_2=-0.8$.
}
\end{figure}

For incoming momenta $\bm{\mathit Q}_1,\bm{\mathit Q}_2$ and exchange momenum $\bm{q}=(0,0)$ or $\bm{q}=(\pi,\pi)$, the interactions in $\mathcal{H}_{\text{int}}$ are shown in Fig.~\ref{FigS3}. We note that all five types of the interactions are positively divergent at the FQ-CAFM phase boundary $K_1^c=-1.6$.

\bibliographystyle{apsrev4-1}
\bibliography{ref}

\begin{thebibliography}{62}
\bibitem [{mis()}]{mis_ed}%
  \BibitemOpen
  \href@noop {} {}\bibinfo {note} {The FQ-CAFM phase boundary is vertical
  ($K_2$~independent) within mean field. Indeed, DMRG calculations on the
  $L_y=6$ cylinder do not show significant $K_2$ dependence of the phase
  boundary for $-1<K_2<-0.4$. For $K_2>-0.4$, the $K_2$ dependence of the
  FQ-CAFM phase boundary becomes visible (increasing $|K_2|$ favors CAFM). Note
  that in DMRG calculations, the FQ-CAFM transition occurs at $K_{1}^c\approx
  -1$, shifted with respect to the mean-field value of -1.6.}\BibitemShut
  {Stop}%
\end{thebibliography}

\end{document}